\documentclass[RNAAS]{aastex62}

\newcommand\bibinc{y}		

\usepackage{subeqnarray}
\usepackage{amsmath}
\usepackage{hyperref}
\usepackage{ulem}

\DeclareMathSymbol{\varOmega}{\mathord}{letters}{"0A}
\DeclareMathSymbol{\varSigma}{\mathord}{letters}{"06}
\DeclareMathSymbol{\varPsi}{\mathord}{letters}{"09}

\begin{document}


\shortauthors{Komacek \& Tan}

\title{Effects of dissociation/recombination on the day-night temperature contrasts of ultra-hot Jupiters}
\author{Thaddeus D. Komacek$^1$ and Xianyu Tan} \affil{Lunar and Planetary Laboratory, University of Arizona, Tucson, AZ, 85721
  tkomacek@lpl.arizona.edu
}

\section{Main body}
\begin{figure}
	\centering	
	\includegraphics[width=0.5\textwidth]{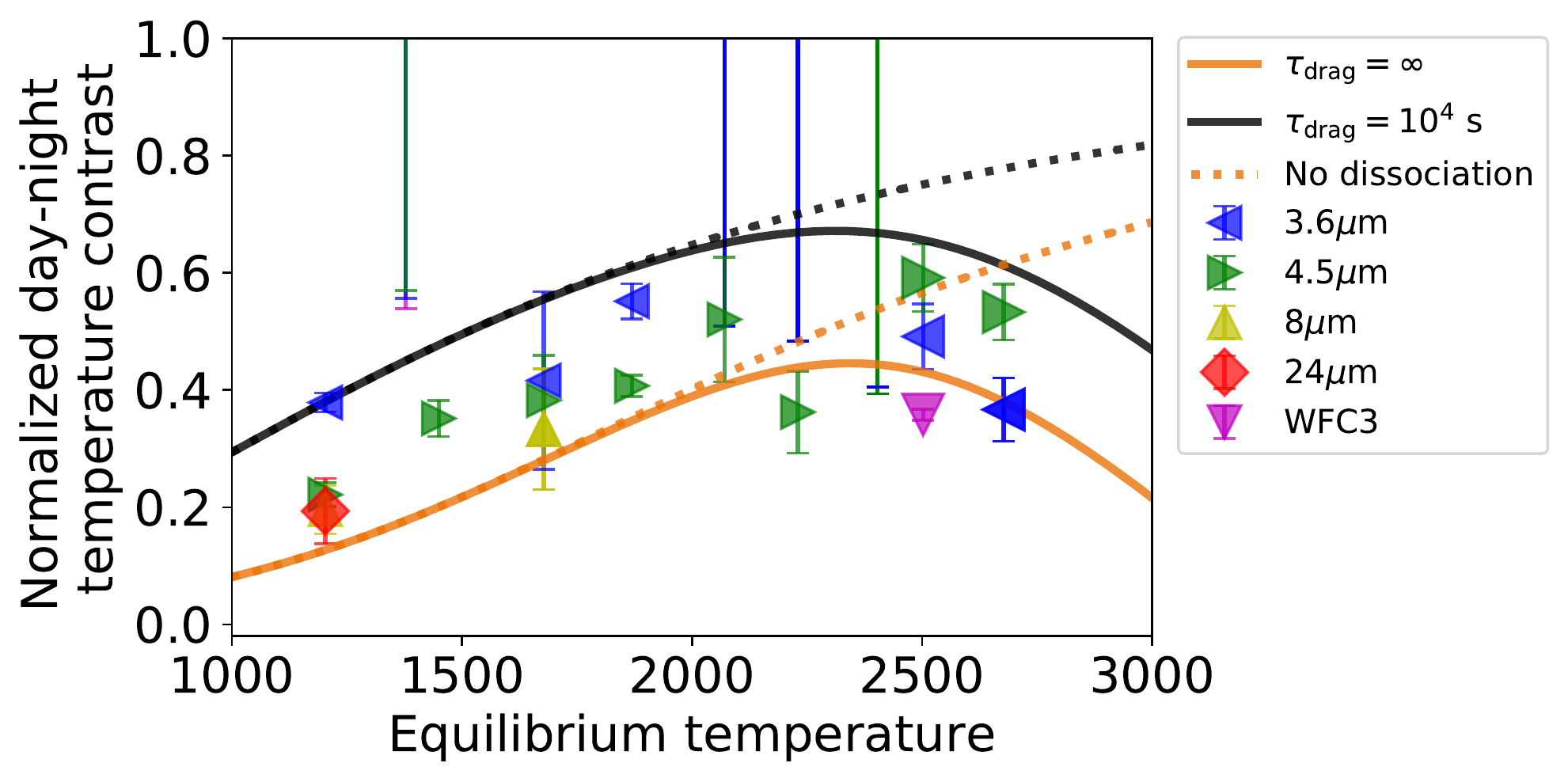}
	\caption{Theoretically predicted dayside-nightside temperature contrast at a pressure of 100 mbars for a planet with the radius, gravity, and rotation rate of HD 209458b (lines, dotted lines do not include dissociation/recombination) and observations (points, lines without a point show lower limits), plotted as a function of equilibrium temperature.}
	\label{fig1}
\end{figure}

\indent Recent observations of ultra-hot Jupiters have 
revealed new physical processes at play that do not occur in the atmospheres of cooler planets. Secondary eclipse observations \citep{Arcangeli:2018,Kreidberg:2018,Mansfield:2018} have found spectra that appear featureless throughout the $1.1-1.7 \mu\mathrm{m}$ region, likely caused by the continuum opacity due to dissociated hydrogen \citep{Bell:2017,Lothringer:2018,Parmentier:2018}. 
Recently, it was shown by \cite{Bell:2018} that when atomic hydrogen is transported from the hot dayside to the relatively cold nightside of these planets, recombination of atomic H into H$_2$ releases a significant amount of heat that can warm up the nightside of the planet. 
Interestingly, recent phase curve observations of the ultra-hot Jupiters WASP-103b \citep{Kreidberg:2018} and WASP-33b \citep{Zhang:2018} showed relatively small amplitudes, indicative of processes reducing their day-night temperature contrast. \\
\indent In this work, we predict the day-night temperature differences of ultra-hot Jupiters using an analytic theory updated from that developed in \cite{Komacek:2015,Zhang:2016,Komacek:2017}. This theory scales the primitive equations of meteorology to solve for characteristic day-night temperature contrasts in hot Jupiter atmospheres.
We include the effects of hydrogen dissociation/recombination, including cooling/heating and the change in mean molecular weight. Specifically, we add the following energy source to the scaled thermodynamic energy equation (Equation 24 in \citealp{Komacek:2015}):  
\begin{equation}
\frac{Q_{\mathrm{recomb}}}{c_p} = \frac{U q_\mathrm{bond} \eta_H(T)}{c_p(T)a} \mathrm{,}
\end{equation}
where $U$ is the wind speed (solved for consistently using the scaled momentum equation), $a$ is the planetary radius, $q_\mathrm{bond} = 2.14 \times 10^8~\mathrm{J}~\mathrm{kg}^{-1}$ is the hydrogen dissociation energy \citep{Bell:2018}, $\eta_H(T)$ is the mass mixing ratio of atomic hydrogen
calculated from the molar mixing ratio $\chi_H$ (given by the approximation in Equation 5 of \citealp{Bell:2018}) as $\eta_H = \chi_H/(2-\chi_H)$, and $c_p(T)$ is the specific heat capacity of the combined hydrogen molecular and atomic gas mixture.
Additionally, hydrogen dissociation decreases the day-side mean molecular weight, which increases the day-night pressure gradient.
We include this effect by modifying the horizontal geopotential difference to take into account the difference between the dayside specific gas constant $R_\mathrm{day}(T) = R_0(1+\eta_H(T))$ and the nightside gas constant $R_\mathrm{night} = R_0$, where $R_0$ is the gas constant of H$_2$. \\
\indent As in \cite{Komacek:2017}, we solve for the dayside-nightside temperature difference as a function of its two key control parameters: incident stellar flux and frictional drag timescale. Figure \ref{fig1} shows our predictions for the day-night brightness temperature difference, $A = (T_\mathrm{b,day}-T_\mathrm{b,night})/T_\mathrm{b,day}$, both including and ignoring the effects of dissociation/recombination. Dissociation and recombination cause the day-night temperature contrast to decrease in the ultra-hot Jupiter regime, in contrast to the case ignoring this effect that shows the opposite trend. Note that for even hotter planets than that considered here (e.g., KELT-9b), the nightside may begin to be partially dissociated, potentially breaking our theoretical assumption that the nightside is constituted of purely molecular hydrogen.
\\ \indent We also show observations of the phase curve amplitude in Figure \ref{fig1}, updated from \cite{Komacek:2017} to include WASP-33b and WASP-103b.
These two phase curve observations do not show evidence for an increase in phase curve amplitude at high temperatures.
We find that hydrogen dissociation/recombination provides a mechanism to explain the relatively small observed phase curve amplitudes of these ultra-hot Jupiters, as predicted by \cite{Bell:2018}.  \\
\indent This theory provides a basic estimate for how dissociation/recombination of hydrogen affects the day-night temperature contrasts of ultra-hot Jupiters as a function of planetary parameters. However, it does not provide details about the feedback of dissociation/recombination on dynamics, 
which is crucial for interpreting observed phase curves. Dissociation of hydrogen on the dayside should cause a strong cooling effect that could affect dynamics near the sub-stellar point. Additionally, because the hydrogen recombination timescales in hot Jupiter atmospheres are extremely short \citep{Parmentier:2018}, the majority of recombination might occur in localized regions with large temperature contrasts (e.g., the limb). As a result, the recombination heating may occur at the limb rather than throughout the nightside as assumed here. Dynamics could then transport heat from the limb toward the nightside of the planet. General Circulation Models that include the effects of hydrogen dissociation/recombination can test our theory, and would provide more detailed comparisons to phase curve observations of ultra-hot Jupiters. \\

We thank Nick Cowan and Taylor Bell for a careful reading of the manuscript and Josh Lothringer for helpful discussions. 

\if\bibinc y

\fi

\end{document}